\documentclass[aps,pra,reprint,groupedaddress]{revtex4-2}
\usepackage{amsmath}
\usepackage{amsfonts}
\usepackage{graphicx}
\usepackage{braket, array, dcolumn}
\usepackage{color}
\begin{document}
\title{Quantum electrodynamics description of localized surface plasmons at a metal nanosphere}
\author{Kuniyuki Miwa}
\email{kuniyukimiwa@ims.ac.jp}
\altaffiliation{Present address: Institute for Molecular Science, Okazaki, Aichi 444-8585, Japan}
\author{George C. Schatz}
\email{g-schatz@northwestern.edu}
\affiliation{Department of Chemistry, Northwestern University, Evanston, Illinois 60208-3113}
\date{\today}
\begin{abstract}
A canonical quantization scheme for localized surface plasmons (LSPs) in a metal nanosphere is presented based on a microscopic model composed of electromagnetic fields, oscillators that describe plasmons, and a reservoir that describes excitations other than plasmons.
The eigenmodes of this fully quantum electrodynamic theory show a spectrum that includes radiative depolarization and broadening, including redshifting from the quasi-static LSP modes, with increasing particle size.
These spectral profiles correctly match those obtained with exact classical electrodynamics (Mie theory).
The present scheme provides the electric fields per plasmon in both near- and far-field regions whereby its utility in the fields of quantum plasmonics and nano-optics is demonstrated.
\end{abstract}
\maketitle
%
Metal nanoparticles (MNPs) have been of great interest in nanotechnology owing to their unique properties originating from localized surface plasmon (LSP) resonances, the collective oscillations of conduction electrons in MNPs~\cite{Maier2007, Kelly2003}.
These resonances exhibit a tremendous potential for manipulating electromagnetic fields beyond the diffraction limit and provide unique control of light, energy, charge, and heat at the nanoscale~\cite{Benz2016, Andrew2004, Li2015, Hsu2017,Brongersma2015}.
A wide range of applications of nanoplasmonics has been reported including nanolasers~\cite{Zhou2013, Ma2019}, optical metamaterials~\cite{Hess2012}, optical nonlinearities~\cite{Kim2008}, photovoltaics~\cite{Atwater2010}, photocatalysis~\cite{Linic2015}, surface-/tip-enhanced Raman spectroscopy~\cite{Nie1997, Kneipp1997}, biosensing~\cite{Anker2009}, and photothermal therapy~\cite{Huang2006}.
Parallel to this prominent progress, the quest for the quantum nature of plasmons and their interaction with matter has triggered a new branch of research named quantum plasmonics~\cite{Tame2013,You2020}.
There have been widespread studies of quantum plasmonics covering such quantum properties as strong coupling~\cite{Chikkaraddy2016}, entanglement~\cite{Altewischer2002}, squeezing~\cite{Huck2009}, and Bose-Einstein condensation~\cite{Hakala2018}.
Quantum plasmonics drives progress in the field of integrated quantum photonics and nano-optics, providing a platform for many technological applications and devices operated at the quantum level, including single-photon sources~\cite{Koenderink2009}, SPASER~\cite{Bergman2003}, transistors~\cite{Chang2007}, ultra-compact circuits~\cite{Heeres2013}, quantum information~\cite{Akimov2007}, and quantum computing devices~\cite{AlonsoCalafell2019}.
\par
The recent upsurge of interest in quantum plasmonics requires a quantum description of both electromagnetic fields and plasmons, which should be described in the natural context of quantum electrodynamics (QED)~\cite{Dirac1927, Craig1998, Salam2009}.
Quantization of electromagnetic fields has been developed since Dirac~\cite{Dirac1927}, however including plasmons with radiative damping and dissipation is a challenge.
A canonical quantization procedure for electromagnetic fields in dispersive and dissipative homogeneous media was proposed by Huttner and Barnett~\cite{Huttner1992}, which is based on prior work by Fano and Hopfield~\cite{Fano1956, Hopfield1958}.
This `microscopic' approach has been extended to several inhomogeneous media subsequently~\cite{Suttorp2004, Bhat2006}.
A different `macroscopic' approach was developed using the Green function formalism and the noise current method~\cite{Gruner1995, Gruner1996}.
The quantum description of plasmons has been developed for bulk materials~\cite{Bohm1953}, metal surfaces~\cite{Elson1971}, and MNPs~\cite{Trugler2008, Waks2010}.
However, both the macroscopic Green function and microscopic Huttner-Barnett approaches have several drawbacks in the quantization process for LSPs in MNPs.
Since the former offers a complicated procedure to calculate the electromagnetic fields, that are obtained indirectly from a phenomenologically introduced noise current operator, it is difficult to physically interpret each mode of the system~\cite{Gruner1995, Gruner1996}.
The latter is based on the Lagrangian formalism and justified in terms of a canonical scheme~\cite{Huttner1992}.
Although this is the most prominent approach, the scheme becomes cumbersome to apply to an inhomogeneous medium.
Therefore, the canonical quantization of LSPs has never been achieved even for a simple metal nanosphere.
The canonical quantization procedure for LSPs was recently presented for a sphere by Shishkov et al~\cite{Shishkov2016}, but only within the quasi-static approximation.
This approximation is valid only for a small particle ($<$ 20~nm) as retardation effects become quite prominent otherwise~\cite{Wokaun1982, Meier1983, Kelly2003}.
\par
A phenomenological approach to quantization of LSPs is widely used and much simpler~\cite{Trugler2008, Waks2010}.
However, in this approach, no canonical formulation is obtained in dispersive and dissipative media.
Moreover, the effects of Joule losses cannot be described in a consistent way~\cite{Shishkov2016}.
For example, eigenfrequencies of LSP resonances are calculated neglecting loss in the quantization procedure~\cite{Tame2013}.
Also, as pointed out in Ref.~\cite{Shishkov2016}, the imaginary part of the permittivity does not affect the electric field generated by LSPs obtained by this approach.
In order to overcome these limitations, it is prerequisite to construct a rigorous approach to the quantization of LSPs which offers a canonical formulation in dispersive and dissipative inhomogeneous media.
\par
In this Letter, we present a fully canonical quantization scheme for LSPs in a dispersive and dissipative metal nanosphere placed in vacuum.
To quantize the electromagnetic fields and plasmons simultaneously, we utilize the Huttner-Barnett model and explore the eigenmodes of the system.
Here, the plasmonic optical response of the metal is modeled with a set of harmonic oscillators that describe linear collective excitations of the electrons~\cite{Fano1956, Hopfield1958}.
In addition, we account for continuum reservoir degrees of freedom (electron-hole pair excitations and phonons) that are coupled to the plasmonic oscillator fields leading to damping~\cite{Huttner1992}.
The reservoir is also responsible for the light absorption, such that diagonalization of the matter part of the Hamiltonian results in a set of dressed continuum fields that describe LSP modes in the quasi-static approximation.
As a second step, the effects of radiation and retardation are investigated by exploring the eigenmodes of the total system composed of the vacuum electromagnetic field and the dressed oscillator field of the matter.
The calculated spectral function correctly exhibits radiation broadening and red-shifting (depolarization) of the plasmon peak due to the light-matter coupling.
By comparing the obtained results with the exact Mie solution from the classical electrodynamics, we find the developed quantum theory can reproduce the exact classical theory well.
Electric fields per plasmon are also calculated and correctly demonstrate both near- and far-field behavior.
Thereby we conclude that the developed theory provides a fully canonical quantization scheme and is valid for both small and relatively large metal nanospheres, including structures where the quasi-static approximation can no longer be applied.
\par
%
We consider a metal nanosphere with radius $R$ composed of  damped harmonic oscillators coupled to vacuum electromagnetic fields.
The Lagrangian is given by
\begin{align}
	L&= \frac{\epsilon_0}{2} \int d^3 \mathbf{r} \left\{
		\left[ \mathbf{\dot{A}}(\mathbf{r}, t) + \nabla \phi(\mathbf{r}, t) \right]^2
		- c^2\left[ \nabla \times \mathbf{A}(\mathbf{r}, t) \right]^2
	\right\}
\nonumber \\
	&+ \frac{\kappa}{2} \int_{r<R} d^3 \mathbf{r} \left\{
		\mathbf{\dot{P}}(\mathbf{r}, t)^2
		-\omega^2_\mathbf{P} \mathbf{P}(\mathbf{r}, t)^2
	\right\}
\nonumber \\
	&+ \frac{1}{2} \int_{r<R} d^3 \mathbf{r} \int^\infty_0 d\Omega \left\{
		\mathbf{\dot{Y}}_{\mathbf{P}\Omega} (\mathbf{r}, t)^2
		-\Omega^2\mathbf{Y}_{\mathbf{P}\Omega} (\mathbf{r}, t)^2
	\right\}
\nonumber \\
	&+ \int_{r<R} d^3 \mathbf{r} \left\{
		\phi(\mathbf{r}, t) \nabla\cdot\mathbf{P}(\mathbf{r}, t)
		+\mathbf{\dot{P}}(\mathbf{r}, t)\cdot\mathbf{A}(\mathbf{r}, t)
	\right\}
\nonumber \\
	&- \int_{r<R} d^3 \mathbf{r} \int^\infty_0 d\Omega
		V_{\mathbf{P}\Omega} \mathbf{P}(\mathbf{r}, t)
		\cdot\dot{\mathbf{Y}}_{\mathbf{P}\Omega} (\mathbf{r}, t),
\end{align}
where $\mathbf{A}(\mathbf{r}, t)$ and $\phi(\mathbf{r}, t)$ represent vector and scalar potential, respectively.
$\epsilon_0$ and $c$ are, respectively, the vacuum permittivity and the speed of light in vacuum, and $t$ is time.
$\mathbf{P}(\mathbf{r}, t)$ indicates a polarization density with the frequency $\omega_\mathbf{P}$ of the harmonic oscillator and the ratio $\kappa$ of the mass to the charge density of the harmonic oscillator.
$\mathbf{Y}_{\mathbf{P}\Omega} (\mathbf{r}, t)$ represents a reservoir composed of a continuum of harmonic oscillators with frequency $\Omega$.
$V_{\mathbf{P}\Omega}$ denotes a polarization-reservoir coupling.
The Lagrangian model is chosen such that it leads to a wave equation which follows from the Maxwell equations in a dissipative and dispersive medium~\cite{Huttner1992}.
The corresponding Lagrangian model has been utilized in Refs.~\cite{Shishkov2016}, where the wave equation in a bulk medium and the bulk permittivity have been derived.
We note that while a single resonance is assumed in this model, the proposed theory can easily be expanded to the many-resonance cases.
It is therefore safe to state that the parameters can be chosen to match the experimentally observed permittivity~\cite{Huttner1992}.
\par
Electromagnetic fields, harmonic oscillator (plasmon) fields, and the reservoir are quantized in a standard manner subject to the commutation rules between the variables and their conjugates~\cite{Craig1998}.
Here, according to the standard approach in nonrelativistic QED, the Coulomb gauge is utilized.
The vector potential $\mathbf{A}$ is expanded onto the vector spherical harmonics and $\phi$, $\mathbf{P}$, and $\mathbf{Y}$ are expanded in terms of the scalar spherical harmonics.
The second-quantized Hamiltonian is given by
\begin{align}
	\hat{H}
	&=\sum_{s=\mathrm{e, o}} \sum_{l=1}^{\infty} \sum_{m=-l}^{l}
		\left[ \hat{h}^{(slm)}_\mathrm{mat} + \hat{h}^{(slm)}_\mathrm{em} \right]
\\
	\hat{h}^{(slm)}_\mathrm{mat}
	&= \hbar\omega_{l} \hat{d}^\dagger_{slm}\hat{d}_{slm}
		+ \int_{0}^{\infty}d\Omega \hbar\Omega \hat{b}^\dagger_{slm\Omega}\hat{b}_{slm\Omega}
\nonumber \\ &~~~~~
		+ \int_{0}^{\infty}d\Omega V_{\mathrm{P}\Omega}\hat{P}_{slm}\hat{\Pi}_{Y_{slm}}(\Omega),
\\
	\hat{h}^{(slm)}_\mathrm{em}
	&= \sum_{\lambda=1}^{2} \int_0^\infty dk \hbar ck
		\hat{a}^\dagger_{\lambda slmk}\hat{a}_{\lambda slmk},
\nonumber \\&~~
	- \int_0^\infty dk
			\frac{\Lambda_{l}(k)}{\kappa R^3l}
			\hat{\Pi}_{P_{slm}} \hat{A}_{2slmk}
\nonumber \\ &~~
		+ \left[ \int_0^\infty dk \frac{\Lambda_{l}(k)}{\sqrt{2\kappa R^3l}}\hat{A}_{2slmk}\right]^2,
\end{align}
with
\begin{align}
	\omega^2_l
	= \omega^2_\mathbf{P}
		+\frac{1}{\epsilon_0\kappa}\frac{l}{2l+1}
		+\frac{1}{\kappa}\int^{\infty}_0 d\Omega V^2_{\mathbf{P}\Omega}.
\label{eq:def-omegal}
\end{align}
Here, $\hat{d}_{slm}$ $(\hat{d}^\dagger_{slm})$ is the annihilation (creation) operator for polarization density with mode $slm$ and frequency $\omega_{l}$, $\hat{b}_{slm\Omega}$ $(\hat{b}^\dagger_{slm\Omega})$ is the reservoir field annihilation (creation) operator for mode $slm$ and frequency $\Omega$, and $\hat{a}_{\lambda slmk}$ $(\hat{a}^\dagger_{\lambda slmk})$ is the annihilation (creation) operator of a transverse photon with wavenumber $k$ and polarization $\lambda$.
$\hat{P}_{slm}$ ($\hat{\Pi}_{P_{slm}}$) and $\hat{Y}_{slm}$ ($\hat{\Pi}_{Y_{slm}}$) are displacement (conjugate momentum) operators for the harmonic oscillator and reservoir fields, respectively.
$\hat{A}_{\lambda slmk}$ is a vector potential operator, and $\Lambda_{l}(k)$ denotes the light-matter coupling strength.
The details of the derivation of the Hamiltonian are shown in the Supplemental Material (SM)~\footnote{See Supplemental Material}.
\par
%
The matter part $\hat{h}^{(slm)}_\mathrm{mat}$ of the Hamiltonian can be diagonalized by the Fano-type of technique~\cite{Fano1961, RosenaudaCosta2000}.
It has been shown that the eigenoperators of this Hamiltonian can be expressed using the permittivity of the bulk medium~\cite{Shishkov2016}.
The spectral function is then obtained by $\rho_\mathrm{qs}(\omega)=-\Im G_\mathrm{qs}(\omega)/\pi$, where $G_\mathrm{qs}(\omega)$ is Fourier transform of the retarded Green function~\cite{Anderson1961, Mahan2000} defined by $G_\mathrm{qs}(t,t') = (1/i\hbar)\theta(t-t')\langle [\hat{d}_{slm}(t), \hat{d}^\dagger_{slm}(t')] \rangle_\mathrm{mat}$ with $\theta(t)$ the step function, $\hat{d}_{slm}(t)$ the operator $\hat{d}_{slm}$ in the Heisenberg representation, $[\cdot,\cdot]$ a commutator of two operators, and $\langle \dots\rangle_\mathrm{mat}$ the statistical average in the representation defined by the system evolution for $\sum_{s,l,m}\hat{h}^{(slm)}_\mathrm{mat}$.
\par
\begin{figure}
\includegraphics[width=8.6 truecm,clip,bb= 0 0 243 230]{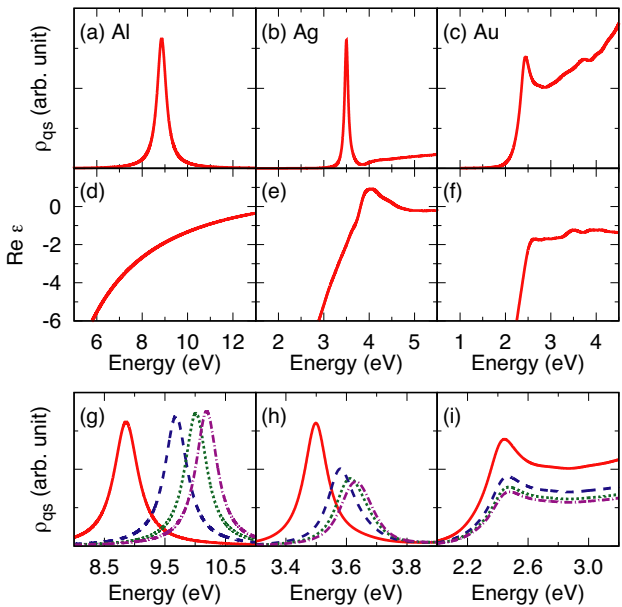}
\caption{
Spectral function $\rho_\mathrm{qs}(\omega)$ for dipolar mode $(l=1)$ for (a) Al, (b) Ag, and (c) Au nanosphere.
Real part of permittivity of (d) Al taken from the experimental data reported in Ref.~\cite{Rakic1998} and (e) Ag and (f) Au in Ref.~\cite{Johnson1972} are displayed.
Spectral function $\rho_\mathrm{qs}(\omega)$ for multipolar modes in (g) Al, (h) Ag, and (i) Au nanosphere are plotted.
Red solid, blue dashed, green dotted, and purple dashed-dotted lines indicate $l=1, 2, 3, 4$ modes, respectively.
\label{fig:QS}
}
\end{figure}
Figures~\ref{fig:QS}(a-c) show the calculated results for $\rho_\mathrm{qs}(\omega)$ for the dipolar mode $(l=1)$.
The spectral profiles are found to be independent of particle radius $R$ as expected for the quasi-static limit.
The peak appears near 8.9~eV for Al, 3.5~eV for Ag, and 2.4~eV for Au, respectively.
The energetic position of the peak in $\rho_\mathrm{qs}(\omega)$ corresponds to the energy satisfying $\Re \epsilon(\omega)=-2$, where $\epsilon(\omega)$ is the metal permittivity [Figs.~\ref{fig:QS}(d-f)].
This relation denotes the so-called Fr\"{o}hlich condition~\cite{Maier2007}, which represents the condition for a resonance excitation of the dipolar LSPs for a small metal nanosphere and is valid within the quasi-static approximation.
It is therefore concluded that the eigenmodes of the matter part of the system provide the LSP modes in the quasi-static approximation.
Figures~\ref{fig:QS}(g-i) demonstrate higher-order multipolar modes $(l>1)$, which reproduce the corresponding LSP modes in the quasi-static approximation.
\par
We now consider the energy region where the imaginary part $\Im\epsilon(\omega)$ of the permittivity is much smaller than $\omega|\partial\Re\epsilon(\omega)/\partial\omega|$.
In this region, the LSP modes can be separated from the reservoir and the eigenmodes of the matter part of the Hamiltonian offer a single discretized mode for each pair of $(s, l, m)$ with the angular frequency $\bar{\omega}_l$, which satisfies $\Re\epsilon(\bar{\omega}_l) = -(l+1)/l$~\cite{Shishkov2016}.
It is shown that such an approximation, which we call the low-loss approximation, is valid in the range $3.0~\mathrm{eV} < \hbar\omega < 15.0~\mathrm{eV}$ for Al, $\hbar\omega < 4.0~\mathrm{eV}$ for Ag, and $\hbar\omega < 2.6~\mathrm{eV}$ for Au~\cite{Note1}.
Since the peak of the LSP modes appears in the energy range where the low-loss approximation is valid, we employ this approximation hereafter.
\par
%
To investigate the retardation and radiation effects, the total Hamiltonian is diagonalized by the Fano-type of technique~\cite{Fano1961, RosenaudaCosta2000}.
The spectral function $\rho_\mathrm{full}(\omega)$ of the LSP modes for the total system is obtained from the imaginary part of the retarded Green function~\cite{Anderson1961, Mahan2000}, where the statistical average is taken in the representation defined by the system evolution for the total Hamiltonian.
The electric field $\mathbf{E}^{(slm)}_{\omega}(\mathbf{r})$ per plasmon is obtained by $\mathbf{E}^{(slm)}_{\omega} (\mathbf{r}) = [\hat{X}^{(slm)}_{\omega}(t), \hat{\mathbf{E}} (\mathbf{r},t) ]$, where $\hat{X}^{(slm)}_{\omega}$ is the eigenoperator for $\hat{h}_{slm}$ with a frequency $\omega$ and the electric field operator $\hat{\mathbf{E}}(\mathbf{r},t)$ is given by $\hat{\mathbf{E}}(\mathbf{r},t) = -\nabla \hat{\phi}(\mathbf{r},t)-\partial\hat{\mathbf{A}}(\mathbf{r},t)/\partial t$ with $\hat{\phi}(\mathbf{r},t)$ and $\hat{\mathbf{A}}(\mathbf{r},t)$ the scalar and vector potential operators, respectively~\cite{Note1}
\par
It can be analytically shown that our theory reproduces the effects of radiation damping and dynamic depolarization on the LSP resonance.
According to the concrete expression of $\rho_\mathrm{full}(\omega)$~\cite{Note1}, the resonance for the dipolar mode ($l=1$) appears at
\begin{align}
	\omega^2 - \bar{\omega}^2_{1}
	 + 36 \bar{\omega}_{1}
		\frac{2\xi^2I(\omega) +i\xi j^2_{1} (\xi)}
			{\left|\partial \Re\epsilon(\bar{\omega}_{1})/\partial\omega \right|}	
	= 0.
\label{eq:RDDD}
\end{align}
with $\xi=\omega R/c$ and $I(\omega)=(R/\pi){\cal P}\int_0^\infty dk j^2_l(kR)/[k^2R^2-\xi^2]$, where ${\cal P}$ denotes the principal part and $j_l(\xi)$ is the spherical Bessel function.
In the following, for the sake of simplicity, we consider the lossless Drude model $\epsilon(\omega)=1-\omega_\mathrm{p}^2/\omega^2$ with $\omega_\mathrm{p}$ the plasmon frequency.
When the value of $\xi$ is small, we obtain $I(\omega) \approx 1/15$ and $j_1(\xi) \approx \xi/3$, and then the resonance condition is given by
\begin{align}
	\left[ \epsilon(\omega)+2 \right]
	- \frac{4}{5} \left[ \epsilon(\omega)-1 \right] \xi^2
	- \frac{2}{3}i \left[ \epsilon(\omega)-1 \right] \xi^3
	= 0.
\end{align}
The second and third terms indicate a shifting and width-broadening of a resonance peak with an increase in $R$, respectively.
The shift and broadening can be scaled as $O(\xi^2)$ and $O(\xi^3)$, respectively.
These phenomena have been intensively investigated in classical electrodynamics~\cite{Wokaun1982, Meier1983, Zeman1984, Kelly2003}, while their analytical derivation based on QED has never been reported as far as we know.
The effects of radiation damping and dynamic depolarization on the LSP resonance are confirmed in numerical calculation results shown below.
\par
\begin{figure}
\includegraphics[clip,bb= 50 50 293 305]{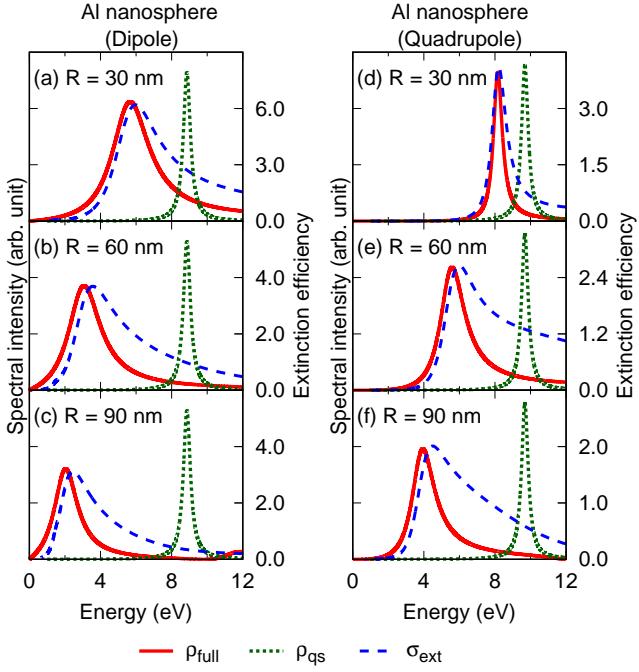}
\caption{
Comparison of the spectral function $\rho_\mathrm{full}(\omega)$ of the total system (red solid line), the spectral function $\rho_\mathrm{qs}(\omega)$ in the quasi-static approximation (green dotted line), and the extinction spectra $\sigma_\mathrm{ext}$ calculated using Mie theory (blue dashed line) for Al nanosphere with radius $R$ of (a)(d) 30~nm, (b)(e) 60~nm, and (c)(f) 90~nm, respectively.
The dipolar (a-c) and quadrupolar modes (d-f) are displayed.
The permittivity $\epsilon(\omega)$ of Al is taken from the experimental data~\cite{Rakic1998}.
\label{fig:Ret_Al}
}
\end{figure}
Figure~\ref{fig:Ret_Al}(a-c) shows the calculated results for $\rho_\mathrm{full}(\omega)$ and $\rho_\mathrm{qs}(\omega)$ for Al nanospheres with different radii $R$ in the case of a dipolar mode.
To confirm the validity of the results, we compare with the extinction spectra $\sigma_\mathrm{ext}(\omega)$ calculated using Mie theory~\cite{Bohren1998}.
The permittivity $\epsilon(\omega)$ of Al is taken from  experimental data~\cite{Rakic1998}.
The results of $\rho_\mathrm{qs}(\omega)$ in the quasi-static approximation exhibit a peak near 8.9~eV independent of $R$.
In $\rho_\mathrm{full}(\omega)$, the peak reflects radiative depolarization and broadening, leading to a peak that is red shifted.
The results are in good agreement with the Mie theory result for $\sigma_\mathrm{ext}(\omega)$.
In addition to the dipolar mode, the quadrupolar mode is considered in Figs.~\ref{fig:Ret_Al} (d-f).
The resonance energy extracted from the peak position in $\sigma_\mathrm{ext}(\omega)$ matches well with the energetic position of the peak in $\rho_\mathrm{full}(\omega)$.
\par
\begin{figure}
\includegraphics[clip,bb= 50 50 293 305]{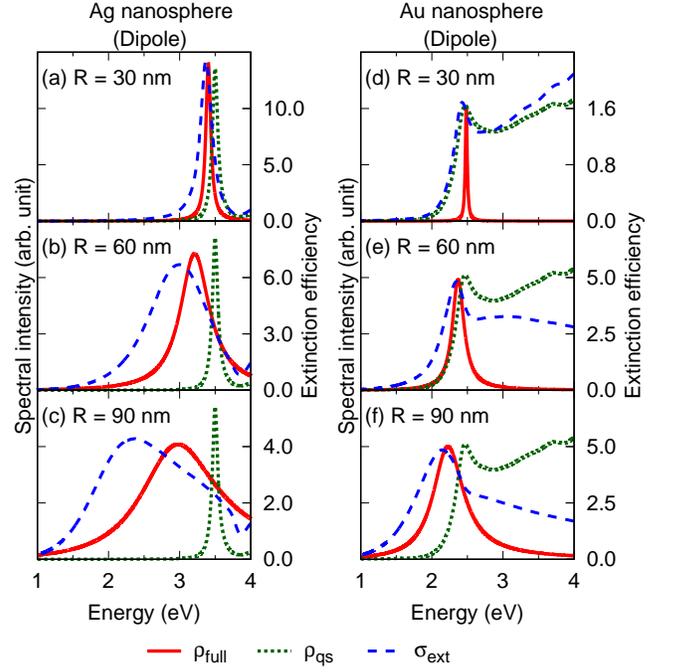}
\caption{
Comparison of $\rho_\mathrm{full}(\omega)$ (red solid line), $\rho_\mathrm{qs}(\omega)$ (green dotted line), and $\sigma_\mathrm{ext}$ (blue dashed line) for Ag and Au nanospheres with radius $R$ of (a)(d) 30~nm, (b)(e) 60~nm, and (c)(f) 90~nm, respectively.
The dipolar mode is displayed.
The permittivity $\epsilon(\omega)$ of both Ag and Au is taken from the experimental data~\cite{Johnson1972}.
\label{fig:Ret_AgAu}
}
\end{figure}
Figure~\ref{fig:Ret_AgAu} exhibits the calculated results for Ag and Au nanospheres obtained using the permittivity $\epsilon(\omega)$ reported in Ref.~\cite{Johnson1972}.
Radiative broadening and peak red shift due to the light-matter coupling are observed.
Owing to the large value of $|\partial\Re\epsilon(\omega)/\partial\omega|$ at $\omega=\bar{\omega}_l$, the linewidth and the amount of the peak shift in $\rho_\mathrm{full}(\omega)$ are narrower and smaller than $\sigma_\mathrm{ext}(\omega)$.
These problems can be solved by introducing a permittivity modeled with the well-known Drude dielectric function~\cite{Zeman1987}.
\par
%
\begin{figure}
\includegraphics[clip,bb= 50 50 293 305]{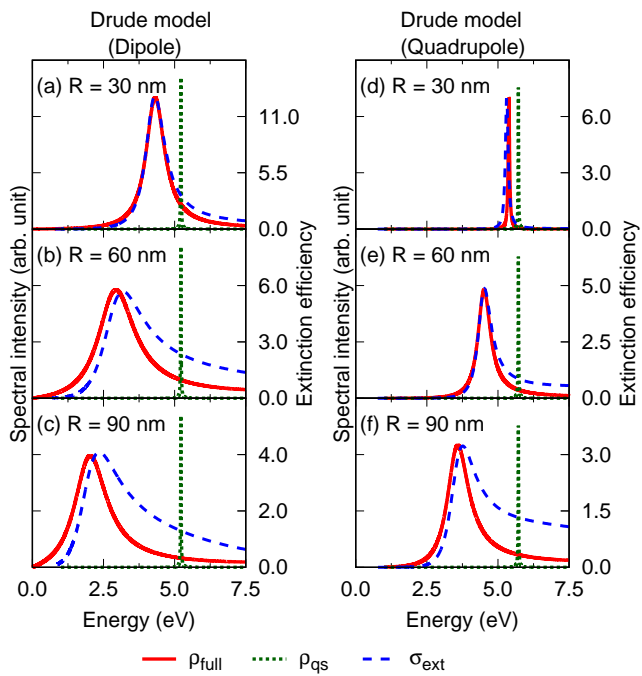}
\caption{
Comparison of $\rho_\mathrm{full}(\omega)$ (red solid line), $\rho_\mathrm{qs}(\omega)$ (green dotted line), and $\sigma_\mathrm{ext}$ (blue dashed line) for metal nanosphere with radius $R$ of (a)(d) 30~nm, (b)(e) 60~nm, and (c)(f) 90~nm, respectively.
The dipolar (a-c) and quadrupolar modes (d-f) are displayed.
The metal permittivity $\epsilon(\omega)$ is parametrized utilizing the Drude model.
The parameters $(\epsilon_{\infty}, \hbar\omega_\mathrm{p}, \hbar\gamma)$ are assumed to be (1.00, 9.04~eV, 21.25~meV) as has been used to model the permittivity of silver~\cite{Zeman1987}.
\label{fig:Ret_Drude}
}
\end{figure}
Figure~\ref{fig:Ret_Drude} shows the calculated results for $\rho_\mathrm{full}(\omega)$ $\rho_\mathrm{qs}(\omega)$, and $\sigma_\mathrm{ext}(\omega)$ for Ag nanoparticles with different radii $R$.
Here, the permittivity of the metal is modeled with the Drude model $\epsilon(\omega) = \epsilon_{\infty} - \omega^2_\mathrm{p}/\omega(\omega+i\gamma)$ with the high-frequency limit $\epsilon_{\infty}$, the plasmon frequency $\omega_\mathrm{p}$, and the damping term $\gamma$.
The parameters $(\epsilon_{\infty}, \hbar\omega_\mathrm{p}, \hbar\gamma)$ are assumed to be (1.00, 9.04~eV, 21.25~meV) as has been used to model the permittivity of silver~\cite{Zeman1987}.
The energetic position of the peak in $\rho_\mathrm{full}(\Omega)$ is in good agreement with the resonance energy extracted from the peak position in $\sigma_\mathrm{ext}(\omega)$.
\par
%
\begin{figure*}
\center
\includegraphics[width=16.0 truecm,clip,bb= 0 0 387 155]{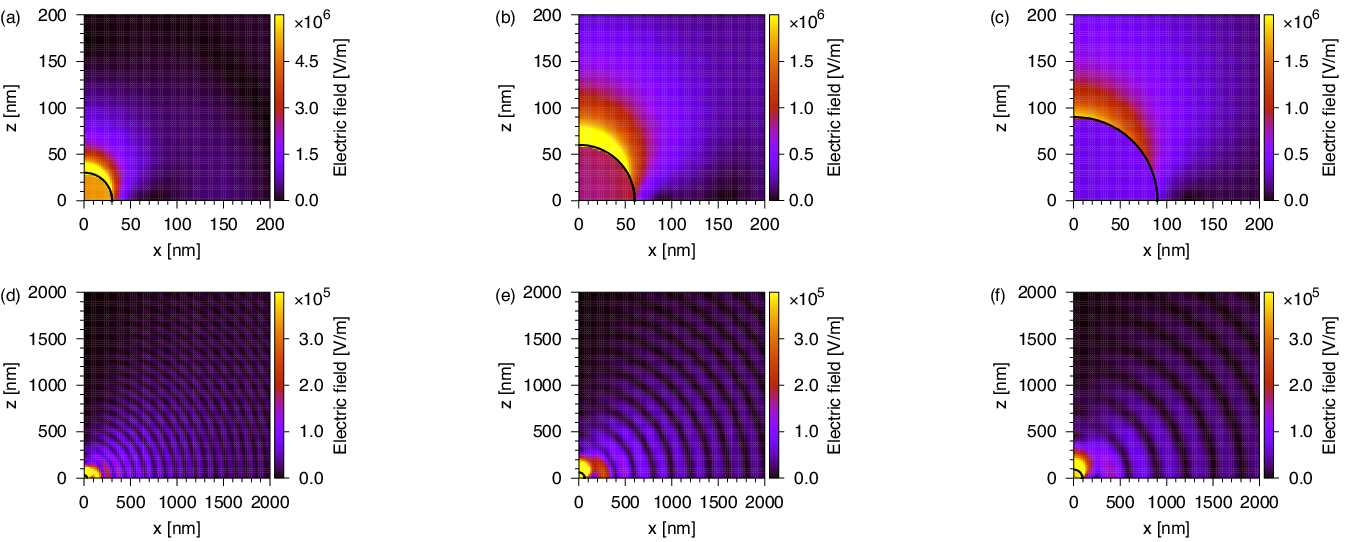}
\caption{
Spatial distribution of electric field $\mathbf{E}^{(slm)}_{\omega}(\mathbf{r})$ per plasmon for Al nanosphere with the radius $R$ for (a,d) 30~nm, (b,e) 60~nm, and (c,f) 90~nm.
Absolute values of the electric fields are plotted.
The angular frequency $\omega$ is set at the resonance excitation energy of (a,d) 5.67~eV, (b,e) 3.08~eV, and (c,f) 2.04~eV.
The permittivity of Al is taken from Ref.~\onlinecite{Rakic1998}.
\label{fig:EField}
}
\end{figure*}
Figure~\ref{fig:EField} presents the calculated results for the electric field $\mathbf{E}^{(slm)}_{\omega}$ per plasmon for Al nanospheres with different radii $R$.
$\mathbf{E}^{(slm)}_{\omega}$ shows both near- and far-field behavior that is familiar for dipolar plasmon excitation~\cite{Bohren1998}.
The intensity of $\mathbf{E}^{(slm)}_{\omega}$ is on the order of 10$^6$~V/m in the near-field region.
The results indicate that, when a quantum emitter (QE) with a dipole moment $\mathbf{d}_\mathrm{QE}$ of several Debye is positioned near a metal nanosphere, the QE-LSP coupling constant $\mathbf{d}_\mathrm{QE} \cdot \mathbf{E}^{(slm)}_{\omega}$ is of the order of several hundred $\mu$eV, which is consistent with the previously reported value~\cite{Bitton2019}.
%
\begin{figure}
\begin{center}
\includegraphics[clip,bb= 50 50 231 220]{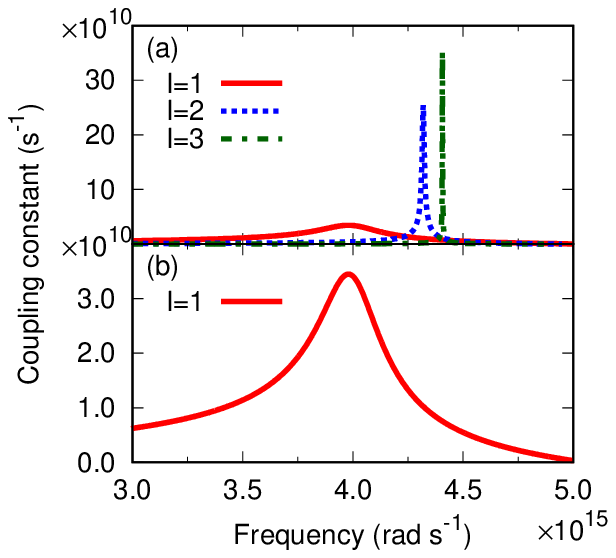}
\end{center}
\caption{
(a, b) Frequency dependence of the coupling constant for the quantum emitter and LSP modes ($l = 1, 2, 3$).
The radius of metal nanosphere is 50~nm and the quantum emitter is located from 5~nm from the metal nanosphere.
The permittivity $\epsilon(\omega)$ of the metal is given by the Drude model with $(\epsilon_\infty, \hbar\omega_\mathrm{p}, \hbar\gamma)$ = (6.00, 7.90~eV, 51.00~meV)~\cite{Varguet2019}.
\label{fig:EFSpectra}
}
\end{figure}
We further note that our theory can reproduce the spectra associated with the coupling constant for a silver nanosphere reported in the previous study~\cite{Varguet2019}.
Figure~\ref{fig:EFSpectra} exhibits the frequency dependence of the QE-LSP coupling constant.
The QE with a dipole moment of 1 Debye is located at 5~nm from a surface of a metal nanosphere with a radius $R = 50$~nm.
The metal permittivity $\epsilon(\omega)$ is given by the Drude model with $(\epsilon_\infty, \hbar\omega_\mathrm{p}, \hbar\gamma)$ = (6.00, 7.90~eV, 51.00~meV).
The results are comparable with those reported in Fig.~8 of Ref.~\cite{Varguet2019}, which indicates the validity of our methods.
\par
It is noteworthy that the evaluation of $\mathbf{E}^{(slm)}_{\omega}$ is crucial for investigating the interaction between the field and quantum emitters, playing an important role in determining Purcell effects associated with the emitter, governing the threshold pumping rate for the lasing oscillations, and many other properties~\cite{Zhou2013, Ma2019, Bergman2003, Noginov2009}.
Moreover, our theory allows the quantitative  analysis of $\mathbf{E}^{(slm)}_{\omega}$ in the far-field region, which provides information on the energy emitted from excited LSPs.
This indicates that our theory provides a useful quantum electrodynamic platform for studying quantum plasmonics and nano-optics.
\par
%
In conclusion, based on a microscopic model for the medium, we developed a fully canonical quantization scheme for the localized surface plasmons (LSPs) associated with a dispersive and absorptive metal nanosphere interacting with the vacuum electromagnetic field.
The matter part of the Hamiltonian is first diagonalized with the Fano technique to determine eigenmodes representing the LSP modes in the quasi-static approximation.
In the energy region where the imaginary part $\Im\epsilon(\omega)$ of the permittivity is much smaller than $\omega|\partial\Re\epsilon(\omega)/\partial\omega|$, the quasi-static LSP modes are shown to be isolated from the reservoir.
Then, using Fano's diagonalization method,  eigenmodes of the total system are obtained, wherein retardation and radiation effects are incorporated into the LSP modes.
The obtained eigenmodes exhibit spectra in which radiation broadening and dynamic depolarization lead to significant broadening and redshifting relative to that of LSPs in the quasi-static approximation.
The energetic position of the peak in the calculated spectra coincides with that obtained from the extinction spectra using Mie theory, which means that the theory matches Maxwell's equations, including all multipoles, for a nanosphere.
The calculated electric fields per plasmon demonstrate realistic behavior in both near- and far-fields, whereby the utility of the developed theory in quantum plasmonics and nano-optics is demonstrated. 
\par
\begin{acknowledgments}
This work was partially supported by National Science Foundation (NSF) under grant CHE-1760537.  Initial work was supported by AFOSR grant FA9550-18-1-0252.
\end{acknowledgments}


\end{document}